\documentclass[twocolumn]{aastex631}  
\usepackage{graphicx,times,epsf}            
\usepackage{longtable}
\usepackage{natbib}
\usepackage{amssymb}
\usepackage{amsmath}
\usepackage{ulem}
\shortauthors{Yang et al.}
\usepackage{soul}

\def\op{\omega_\mathrm{p}}
\def\os{\omega_\star}

\begin{document}

\title{Transit Timing Variation of XO-3b: Evidence for Tidal Evolution of Hot Jupiter with High Eccentricity}
\email{Fan Yang: sailoryf@nao.cas.cn; sailoryf1222@gmail.com}
\email{Xing Wei: xingwei@bnu.edu.cn}

\author[0000-0002-6039-8212]{fan yang}
\affil{Department of Astronomy, Beijing Normal University, Beijing 100875,
People's Republic of China}
\affil{National Astronomical Observatories, Chinese Academy of Sciences, 20A
Datun Road, Chaoyang District, Beijing 100101, China\\}
\affil{School of Astronomy and Space Science, University of Chinese Academy of Sciences,
Beijing 100049, China\\}

\author{Xing Wei}
\affil{Department of Astronomy, Beijing Normal University, Beijing 100875,
People's Republic of China}

\begin{abstract}
Observed transit timing variation (TTV) potentially reveals the period decay caused by star-planet tidal interaction which can explain the orbital migration of hot Jupiters. We report the TTV of XO-3b, using TESS observed timings and archival timings. We generate a photometric pipeline to produce light curves from raw TESS images and find the difference between our pipeline and TESS PDC is negligible for timing analysis. TESS timing presents a shift of 17.6 minutes (80 $\sigma$), earlier than the prediction from the previous ephemeris. The best linear fit for all timings available gives a Bayesian Information Criterion (BIC) value of 439. A quadratic function is a better model with a BIC of 56. The period derivative obtained from a quadratic function is -6.2$\times$10$^{-9}$$\pm$2.9$\times$10$^{-10}$ per orbit, indicating an orbital decay timescale 1.4 Myr. We find that the orbital period decay can be well explained by tidal interaction. The `modified tidal quality factor' $Q_{p}'$ would be 1.8$\times$10$^{4}$$\pm$8$\times$10$^{2}$ if we assume the decay is due to the tide in the planet; whereas $Q_{*}'$ would be 1.5$\times$10$^{5}$$\pm$6$\times$10$^{3}$ if tidal dissipation is predominantly in the star. The precession model is another possible origin to explain the observed TTVs. We note that the follow-up observations of occultation timing and radial velocity monitoring are needed for fully discriminating the different models. 
\end{abstract}
\keywords{Exoplanet systems (484), Exoplanet astronomy (486), Transit photometry(1709), transit timing variation method (1710)}

\section{Introduction}

Transit Timing Variation (TTV) potentially provides direct observational evidence of tidal migration which probably explains the origin of hot Jupiter, holding two processes, i.e., the reducing of the momentum and then the energy \citep[][and reference therein]{Hansen2010, Li2010, Dawson2018}. The orbital momentum reduces when the eccentricity ($e$) is excited by the perturbations, e.g., planet scattering, secular interaction \citep{Juric2008, Wu2011, Endl2014, Petrovich2015, Shara2016}. The tides dissipate the orbital energy which turns into the heat within the host star and Jupiter \citep{Fabrycky2007, Naoz2011, Penev2014}.

Transiting Exoplanet Survey Satellite \citep[TESS;][]{Ricker2015, Fausnaugh2021} launched in 2018, provides high precision photometry for hot Jupiter transits in the whole sky. Combining the most recent TESS result and archival data, it is capable to monitor TTVs in need of tidal dissipation analysis. WASP-12b has been reported with orbital decaying through TTV monitoring in recent decade, including the newest results from TESS  \citep{2011wasp12b,2012wasp12b,2017wasp12b,2020wasp12b, 2021wasp12bTESS}. WASP-4b shows an earlier timing than ephemeris prediction at 81.6$\pm$11.7 s which turns out to be due to the R$\o$mer effect according to the further investigation \citep{WASP-4b,wasp-4b2020}.

TTVs potentially answer the key question that if the planet migrates to the close-orbit at the early formation due to the disk friction, or at random stages due to the perturbation and tidal dissipation \citep{Lin1996, Wu2003, Dawson2018}. The latter scenario 
would be either predominant or an important supplement to the hot Jupiter migration, when hot Jupiters are discovered with a period-decaying timescale significantly shorter than star-planet age, especially if the orbital eccentricity is high. 

XO-3b is reported as a migration candidate inferred from holding a high $e$ of 0.27587$^{+0.00071}_{-0.00067}$ and high $M_{p}$/$M_{\ast}$ of 9$\times$10$^{-3}$ \citep{Bonomo2017}. \citet{Hebrard2008} obtains an obliquity of 70$\pm$15 degrees from the Rossiter-McLaughlin effect. These static system properties are believed to indicate XO-3b should hold a period decay process and should present observable TTVs.  
Interestingly, XO-3b is also reported as a candidate of apsidal precession \citep{Jordan2008, Nascimbeni2021} which could be another possible physical contributor of observed TTVs.

In a previous work \citep[][hereafter Paper I]{shan2021}, we build a sample of 31 hot Jupiters with significant TESS timing offsets to the previous ephemeris, using TESS Objects of Interest (TOI) Catalog \citep{TOIcatalog}. Among the sample, XO-3b is one of the most significant offset sources and is briefly discussed as a candidate holding period decay.

In this work, we obtain timings of XO-3b from TESS light curves, present TTV evidence by combining the archival timings for more than ten years, and investigate possible TTV origins. The paper is organized as follows. In section 2,  we describe the data reduction and transit fitting. 
In section 3, we present the XO-3b TTV and the most likely explanation. In Section 4, we give a brief summary and discussion.

\section{data reduction and transit modeling}

XO-3b is a hot Jupiter transiting a F5V-type star with an orbital period of 3.19 days \citep{Winn2008}. The planet has a mass of 11.70$\pm$0.42 $M_{J}$, a radius of 1.217$\pm$0.073 $R_{J}$. The host star has a mass of 1.213$\pm$0.066 $M_{\odot}$, a radius of 1.377$\pm$0.083 $R_{\odot}$ \citep{Bonomo2017}. The planet-star system has a semi-major axis of 4.95$\pm$0.18 \citep[in stellar radii][]{Stassun2017}. 

\subsection{TESS Light Curve}

\begin{figure}
  \centering
  \includegraphics[width=3.5in]{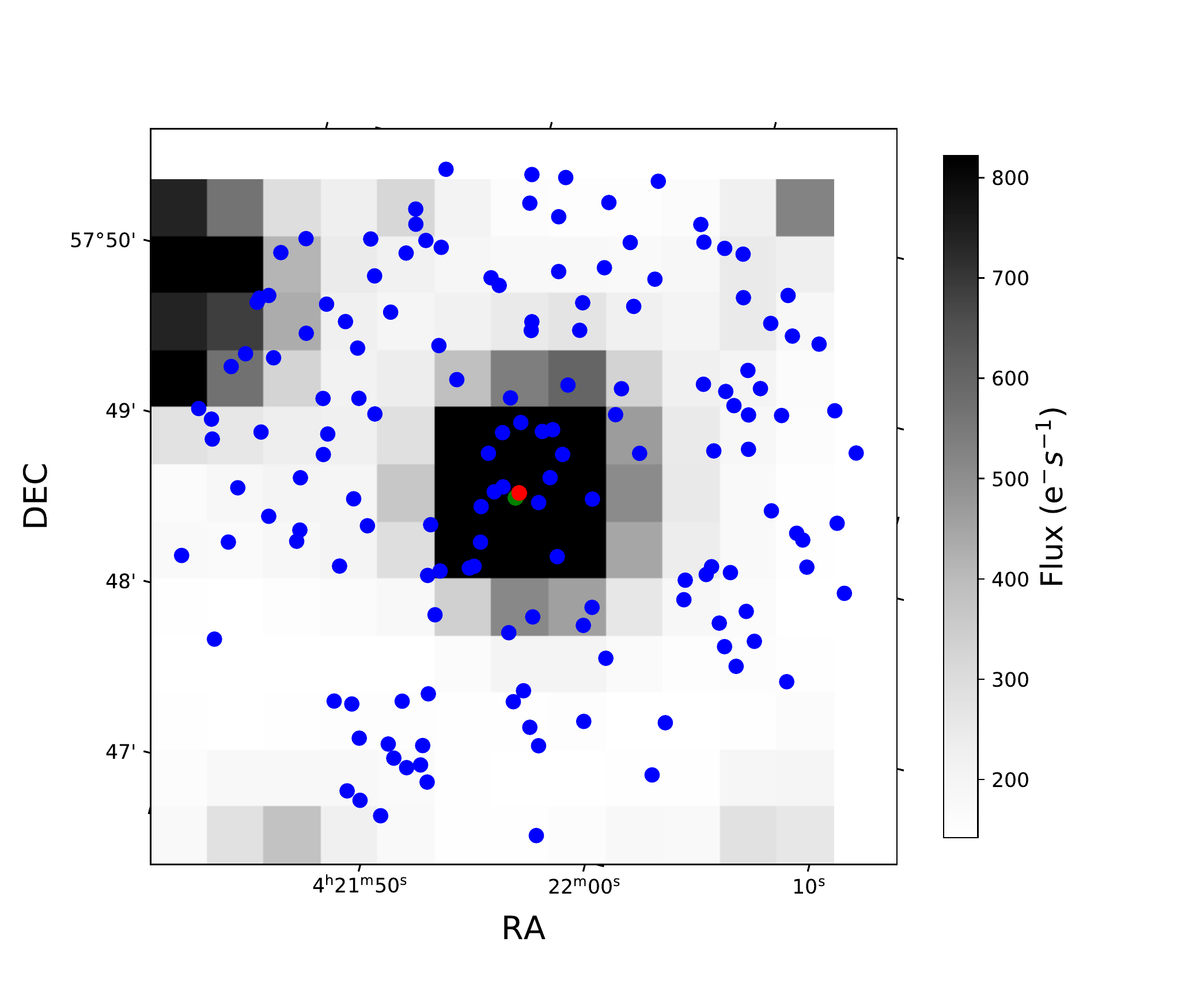}
  \caption{TESS cut-off image of XO-3b at 14$\times$14 pixels. The red point indicates the target position after astrometry correction, while the green point gives the uncorrected target position in the Gaia catalog \citep{GaiaDr2}. The blue points present the nearby source positions within a radius of 6 TESS pixels (126 arcseconds).}
  \label{image:frame}
\end{figure}

TESS processes a field of view (FOV) of 24×96 deg$^{2}$ and a pixel resolution of 21 arcseconds \citep{Ricker2015}. TESS provides continuous photometry for the same sky area in 27 days (a sector). The images are co-added to 30 minute cadence for the full frame images (FFI) and to 2 minute cadence near targets of interest (target pixel files; TPF). The recently released FFIs apply 10 minute cadence after the first two-year sky survey. An example image of XO-3b is present in Figure \ref{image:frame}. High-quality TESS photometry has been well utilized in exoplanet researches, e.g., the atmospheric measurement, transit timing variation \citep{Deming2009, Placek2016, Kempton2018, Yangatmos, wasp-4b2020, YangLD, Kane2021, 2021wasp12bTESS}.

We generate a photometry pipeline, to obtain light curves from TESS raw images. The pipeline includes the modules of e.g., astrometry correction, contamination deblending, circular aperture photometry, and light curve detrending \citep[details as described in][]{Yangatmos}. The pipeline generates a consistent result when compared to Pre-search Data Conditioning (PDC) result \citep{Stumpe2014} from the Science Processing Operations Center (SPOC) pipeline, among the recent applications \citep{Yangatmos, YangLD, Yanghats5b}. The light curve generated by our pipeline enables us to obtain the transit parameters of the sources without PDC light curves \citep[similar to another TESS photometry pipeline;][]{Feinstein2019}.

XO-3b is observed by TESS from 2019-11-28 to 2019-12-23 and is available with the PDC light curve. We generate light curves from TPF 2-minute images and find the difference of the conjunction timings between our result and PDC result is within 0.3 $\sigma$ (0.1 minutes). In this work, we apply the result from PDC light curve, in the convenience of the potential follow-up comparison.

A detrending process is needed to remove the systematics in the light curve, including the residual instrumental effect
and the potential physical effect such as stellar variability. These systematics can be removed by a simple linear detrending when analyzing the transit timings \citep[details in][]{Yangatmos, YangLD}. TESS continuous observation enables one to well model the baseline near the transit. A linear function is applied when fitting the baseline. We also try higher-order polynomial functions which give a negligible difference to the transit timing.

\begin{figure*}
  \centering
  \includegraphics[width=8in]{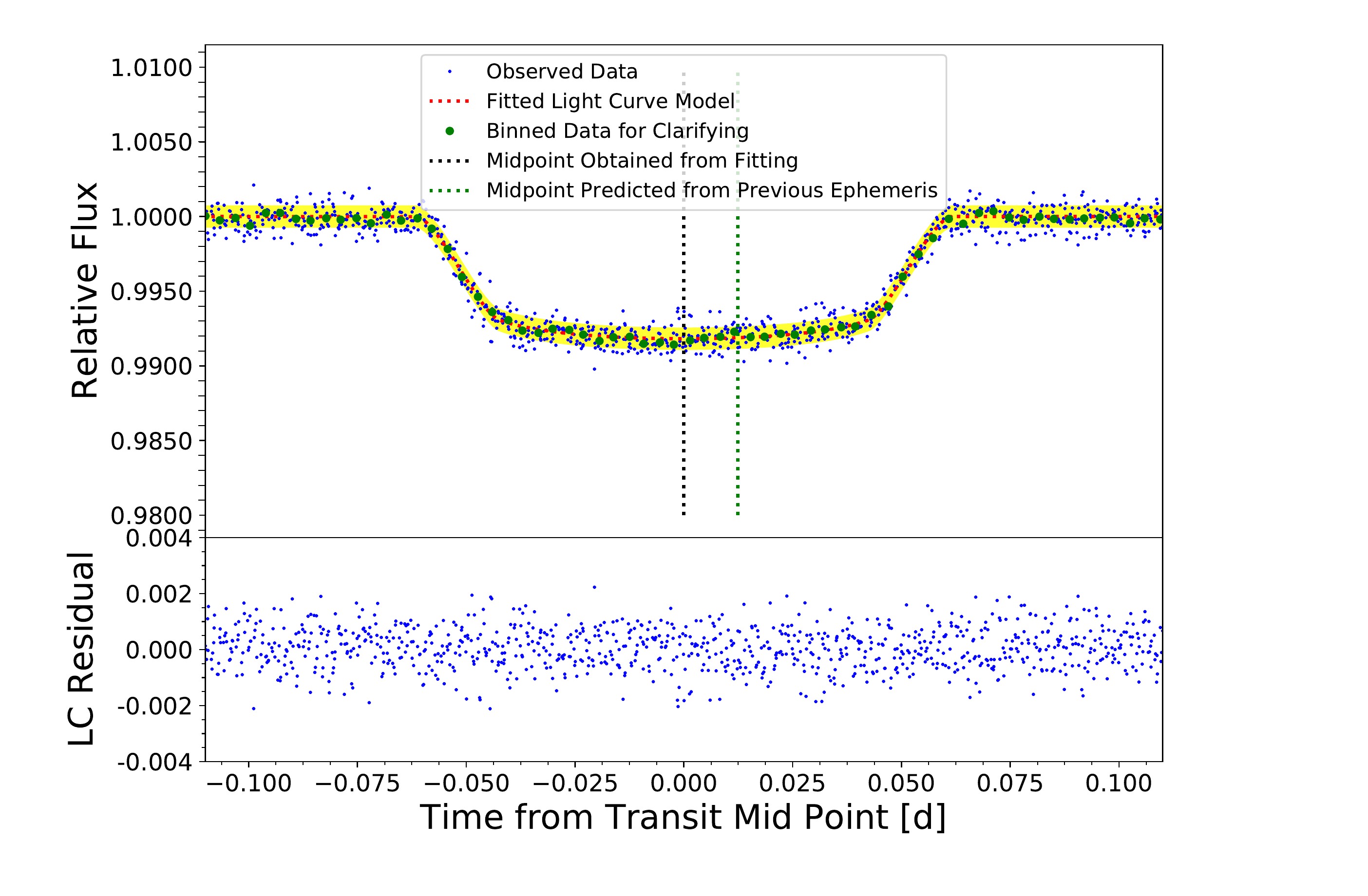}
  \caption{Folded light curve of XO-3b. The blue points present TESS observations; green points, binned data sets for clarity; red line, fitted light curve model; yellow region, 1 $\sigma$ significant region; black vertical line, conjunction TESS timing obtained from modeling the light curve; green vertical line, the predicted timing from previous ephemeris \citep{Wong2014}. The previous ephemeris prediction is 17.8 minutes later than TESS timing. The bottom panel gives the fitting residual, showing no structure. The residual standard deviation is 751ppm.}
  \label{image:lc}
\end{figure*}

\subsection{Transit Modeling}
\label{sec:fit}
The transit timing is obtained by modelling the light curve \citep{Mandel_Agol2002}, using Markov chain Monte Carlo (MCMC) \citep{pymc, emcee}. The MCMC chain takes 80,000 steps after the first 30,000 steps as burn-in. We fit the light curve with a `Keplerian orbit', being consistent with the reported eccentricity as high as 0.27587$^{+0.00071}_{-0.00067}$ \citep{Bonomo2017}. The inclination ($i$) is fixed to 79.32, according to the reference work \citep{Stassun2017}.

The free parameters include transit mid-point, the semi-major axis ($a/R_{\ast}$), the radius ratio of the planet to the host star ($R_{p}/R_{\ast}$), the quadratic limb darkening coefficients (u1 and u2), the argument of periapsis, the longitude of the ascending node, the time of periapse passage. The priors of the free parameters are all uniform except for the limb darkening. The limb darkening prior is applied as a Gaussian distribution with a $\sigma$ of 0.05 \citep{Yangatmos}, centering at the stellar model prediction \citep{TESSLD}. The limb darkening prior is 0.32$\pm$0.05 (u1), 0.22$\pm$0.05 (u2), interpolated based on the stellar parameters from TESS input catalog \citep{Stassun2019}.

We fit light curves of every transit visit as well as the conjunction light curve folded by the previous ephemeris \citep{Wong2014}. The center of conjunction light curve among one TESS sector ($\sim$ 30 days) would be shifted at $\sim$ 1 minute if the folding period showing any difference at 0.0002 days. The period precision we discuss in this work is at the level of 0.00001 days which would not induce any significant timing uncertainty to the conjunction transit mid-point. This fold-and-check algorithm has been commonly used in the analysis of binary researches \citep{Yang2020, PanFu2021}.

\section{TTV evidence and physical explanations}
\subsection{Timing Analysis}

TESS timings are obtained through modeling the light curves (as shown in Figure \ref{image:lc}). We list the timings for every single transit visit and the conjunction timing in the whole TESS sector (in Table \ref{table: timings}). Single transit timings and conjunction timing are well consistent within 1 $\sigma$. We note that the difference between HJD and BJD is within 4s which is beyond the timing precision discussed in this work. We thereby do not discriminate between HJD and BJD in timing analysis.

\begin{table*}
\setlength{\tabcolsep}{3mm}
\begin{center}
\caption{XO-3b TESS timings}
\label{table: timings}
\begin{tabular}{cccc}
  \hline
 \hline
\multicolumn{4}{c} {TESS Transit Mid-points (HJD-2457000)} \\
\hline    
1819.06428$\pm$0.00035 &   1822.25556$\pm$0.00034 &   1825.44732$\pm$0.00037 &   1831.83008$\pm$0.00035 \\
1835.02191$\pm$0.00034 &   1838.21397$\pm$0.00042  & 1831.83039$\pm$0.00016$^a$  & 1819.064098$\pm$0.00028$^b$   \\
\hline
\multicolumn{4}{c} {Transit Mid-point Variation with Respect to previous Ephemeris$^a$ [minute]} \\
\hline   
-17.54$\pm$0.50 &   -17.91$\pm$0.49 &   -17.57$\pm$0.53 &   -18.02$\pm$0.51 \\   -17.59$\pm$0.48 &   -16.83$\pm$0.61 &  -17.57$\pm$0.22$^a$  &  -17.80$\pm$0.40$^b$ \\
\hline 
\end{tabular}
\end{center}
\begin{flushleft}
Note.
(a) Conjunction timing folded on the period from \citet{Wong2014}. 
(b) Timing from \citet{TOIcatalog}.
\end{flushleft}
\end{table*}

\begin{figure*}
  \centering
  \includegraphics[width=8in]{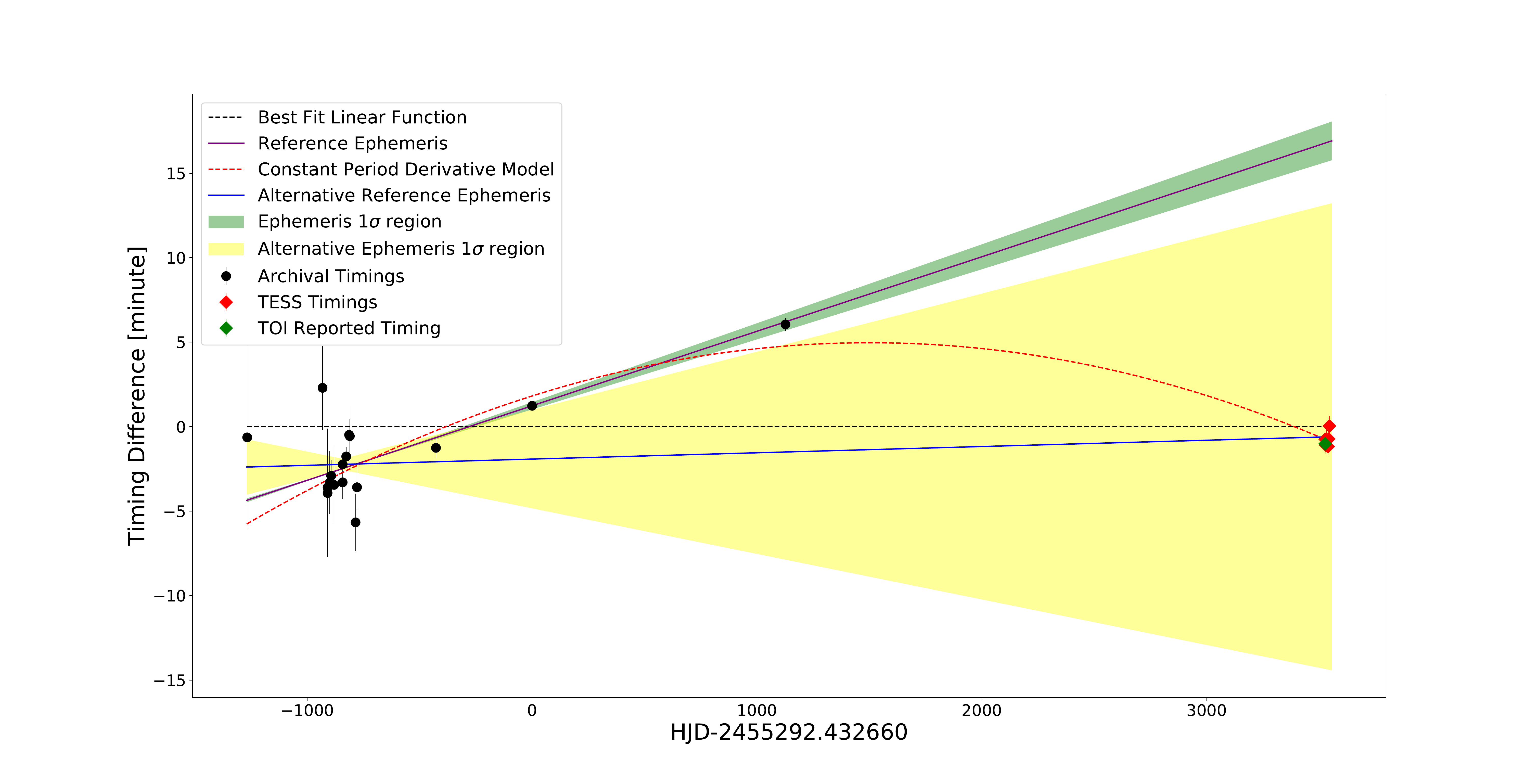}
  \caption{Evolution of XO-3b timings. The baseline (black dash line) is set as the linear fit to all the available timings. Black points show the archival timings; red diamonds, TESS timings of transit visits; green diamond, TOI reported conjunction timing. The purple solid line and green region present the ephemeris predicted median timing and 1 $\sigma$ region from \citep{Wong2014}; 
the blue solid line and yellow region, the ephemeris prediction from \citep{Winn2008}. The red line gives the best fitted quadratic function, indicating a constant period derivative model.}
  \label{image:timings}
\end{figure*}

The timings (as shown in Figure \ref{image:timings}) show a significant inconsistency at the recent TESS observations when compared to the most recent reported ephemeris \citep{Wong2014}. TESS conjunction timing is 17.6 minutes ($\sim$ 80 $\sigma$) earlier than the ephemeris prediction. Any linear function, indicating a constant period model \citep[][and references therein]{Winn2008, Winn2009, Johns-Krull2008, Wong2014, Bonomo2017}, does not well fit the observational timings. The Bayesian Information Criterion (BIC) value of the best linear fit is 439 \citep[BIC details as described in][]{BICmostcited}. Conversely, the BIC is 56 for a quadratic function which means a constant period derivative model. The physical interpretation of such a period derivative model is available in Section \ref{sec:explanation}.
The $\Delta$(BIC) is 383, indicating that the quadratic model is significantly preferred (as shown in Table \ref{table: models}).

Bayesian analysis is sensitive to the data uncertainties.
Any refinement of timing uncertainties could significantly modify the BIC values. We test the robustness of BIC preference by amplifying the timing uncertainties by factors of 2 and 3. The $\Delta$(BIC) is 98 when timing uncertainties are enlarged twofold. The $\Delta$(BIC) is 47 when uncertainties increase threefold.

The accuracy of our generated TESS timing is evaluated by comparing it to the timing reported by TESS Objects of Interest (TOI) Catalog \citep{TOIcatalog}. TOI timing of XO-3b is highly consistent with our generated TESS timings. The difference between the TOI timing and our conjunction timing (when folded into the same epoch) is within 0.3 minutes (0.5 $\sigma$). In Paper I, we compare the timings of 262 hot Jupiters listed in the TOI catalog to the archival ephemeris \citep{ExoplanetArchive} and find 159 of them show consistent results. We check our generated timings and TOI timings, finding no significant difference (details in Paper I). XO-3b is one of three sources presenting the largest timing offsets among which WASP-161b has been reported with TTV evidence (Yang et al. submitted). In reference works, we also present the comparison between our generated TESS timings and previous ephemeris. The comparison sample includes WASP-58b and HAT-P-31b, presenting a negligible difference.

\subsection{Possible Physical Explanations}
\label{sec:explanation}

The star-planet tidal interaction induces the torque to transfer angular momentum between orbit and spins and the energy dissipation inside star and planet, such that it influences the orbital dynamics in some respects: migration (the change in semi-major axis or equivalently orbital period), synchronization (the change in the difference of orbital and spin frequencies), circularization (the change in orbital eccentricity), and the change in orbital obliquity \citep{Goldreich1966, Hut1981, Bodenheimer2001, Ogilvie2014}. The time scale of circulization ($\sim$ 1 Myr) is much smaller than that of host star evolution. In addition, high eccentricity can be excited by some other mechanisms, e.g., planet-planet scattering, Kozai-Lidov cycles, secular chaos \citep{Wu2011, Naoz2011, Dawson2018}. High eccentricity can induce strong tidal dissipation which in turn induces orbital decay, that is the most plausible explanation for the observed TTV.

The period decay model predicts a quadratic function between the transit times and the Nth transit epoch \citep{2017wasp12b}: 
\begin{equation*}
\begin{aligned}
t_\mathrm{tra}(N) &= t_0 + NP + \frac{1}{2}\frac{dP}{dN}N^2.\\
\end{aligned}
\end{equation*}
$t_\mathrm{tra}(N)$ is the Nth transit timing, $t_0$ is the timing zero point. Therefore, the TTV is a function of time as well (as shown Table \ref{table: models}). Fitting the TTVs, the derived period derivative ($\dot{P}$) is -6.2$\times$10$^{-9}$$\pm$2.9$\times$10$^{-10}$ days per orbit per day. The orbital decay timescale $P$/$\dot{P}$ is 1.4 Myr.

\begin{table}
\setlength{\tabcolsep}{3mm}
\begin{center}
\caption{Models for Observed TTV}
\label{table: models}
\begin{tabular}{cc}
  \hline
 \hline
\multicolumn{2}{c} {Linear Model: t$_{d}$ = a$_{l}$$\times$t + b$_{l}$} \\
\multicolumn{2}{c} {Period Decay Model: t$_{d}$ = a$_{d}$$\times$t$^{2}$ + b$_{d}$$\times$t + c$_{d}$} \\
\hline   

Parameters  &  Timing   \\
 \hline
                    \multicolumn{2}{c}   { Linear Model }\\
\hline 
 
a$_{l}$ &      -0.00441$\pm$6.1$\times$ 10$^{-5}$  \\
b$_{l}$ &      -1.23$\pm$0.13  \\
BIC   &  439 \\
 \hline
\multicolumn{2}{c} {Period Decay Model} \\
\hline 

a$_{d}$     &   -1.396$\times$10$^{-6}$$\pm$6.6$\times$10$^{-8}$ \\
b$_{d}$     &    -0.0002$\pm$0.0002 \\
c$_{d}$     &    0.58$\pm$0.16 \\
$\dot{P}$ (per orbit) & -6.2$\times$10$^{-9}$$\pm$2.9$\times$10$^{-10}$\\
BIC   &  56 \\
\hline 
\end{tabular}
\end{center}
\begin{flushleft}
\end{flushleft}
\end{table}

The rate of orbital period change $\dot{P}$ can be estimated with the theory of equilibrium tide \citep{Hut1981,Themodel},
\begin{equation}
\label{modelapplied}
\begin{split}
\dot{P} = \frac{27\pi}{Q_p'}\left(\frac{M_\star}{M_p}\right)\left(\frac{R_p}{a}\right)^5\left[N(e)\,x_\mathrm{p}\,
\frac{\op}{n}-N_a(e)\right] \\
+ \frac{27\pi}{Q_*'}\left(\frac{M_p}{M_\star}\right)\left(\frac{R_\star}{a}\right)^5\left[N(e)\,x_\star\,
\frac{\os}{n}-N_a(e)\right],\\
\end{split}
\end{equation}
where the subscript `*' stands for the stellar parameters and `p' for the planet parameters, $M$ denotes mass, $R$ radius, $a$ semi-major axis, $n$ mean motion, $\omega$ rotation rate, $x = \cos \epsilon$, where $\epsilon$ is the obliquity, and $Q'$ modified tidal quality factor.
$N(e)$ and $N_a(e)$ are determined by eccentricity,
\begin{equation*}
\label{n_e}
N(e) = \frac{1+\frac{15}{2}e^2+\frac{45}{8}e^4+\frac{5}{16}e^6}{(1-e^2)^{6}}
\end{equation*}
and
\begin{equation*}
\label{na_e}
N_a(e)=\frac{1+\frac{31}{2}e^2+\frac{255}{8}e^4+\frac{185}{16}e^6+\frac{25}{64}e^8}{(1-e^2)^{15/2}}.
\end{equation*}

According to Equation \eqref{modelapplied}, $Q'_\star$ can be estimated to be 1.5$\times$10$^{5}$$\pm$6$\times$10$^{3}$ with the assumption that the orbital decay arises from the tide raised by planet on star. $Q'_p$ can be estimated to be 1.8$\times$10$^{4}$$\pm$8$\times$10$^{2}$ if the orbital decay arises from the tide raised by star on planet. The estimation is based on these parameters: stellar mass $M_*=1.213\pm0.066 M_\odot$, stellar radius $R_*=1.377\pm0.083 R_\odot$, planet mass $M_p=11.70\pm0.42M_J$, planet radius $R_p=1.217\pm0.073 R_{J}$, semi-major axis $a=4.95\pm0.18 R_*$, eccentricity  $e=0.27587^{+0.00071}_{-0.00067}$, and obliquity $\epsilon=70\pm15^\circ$ for the star and zero for the planet \citep{Hebrard2008,Bonomo2017,Stassun2017}. We note that applying certain parameters can enlarge the obtained value of $Q'_\star$ and $Q'_p$, e.g., applying stellar obliquity of 37.3$\pm$3.7 \citep{Winn2009} would double $Q'_\star$ value.

The system parameters imply the history of eccentricity excitation. \citet{Bonomo2017} reports the stellar age 2.82$^{+0.82}_{-0.58}$ Gyr and the projected rotational velocity 18.54$\pm$0.17 km s$^{-1}$. At such an age, the orbit should have already been circularized. Moreover, a pseudo-synchronization between stellar spin and the planet revolution has achieved, which is indicated by the stellar projected rotational velocity and planet orbital period \citep{Hut1981,Levrard2009,Bonomo2017}. We thereby set values of $\frac{\op}{n}$ and $\frac{\os}{n}$ applied in Equation \ref{modelapplied} as both 1. The large eccentricity (0.27587$^{+0.00071}_{-0.00067}$) is thus probably due to the recent (less than a few Myr) perturbations \citep{Dawson2018}.

XO-3b has been reported to experience tidal dissipation through the static systematic features \citep{Bonomo2017}, e.g., its large eccentricity and high mass ratio of the host star to the planet.
Also, these static features imply that XO-3b would hold precession \citep{Jordan2008, Nascimbeni2021}, especially when the obliquity is as high as 70$\pm$15 degrees \citep{Hebrard2008}. It is worth investigating the contribution of the precession towards TTV.

Apsidal precession commonly affects the light curve shape more significantly than the timing; and the transit duration variation (TDV) is a more evident observational parameter than TTV for the precession measurement \citep{Pal2008, Ragozzine2009}, except for the case when the impact parameter b closes to 1/$\sqrt{2}$ \citep[Equation 15 in][]{Jordan2008}. We compare the transit durations of TESS light curves and light curves from \citep{Winn2008}, producing a TDV baseline longer than 10 years.

\begin{figure*}
  \centering
  \includegraphics[width=8in]{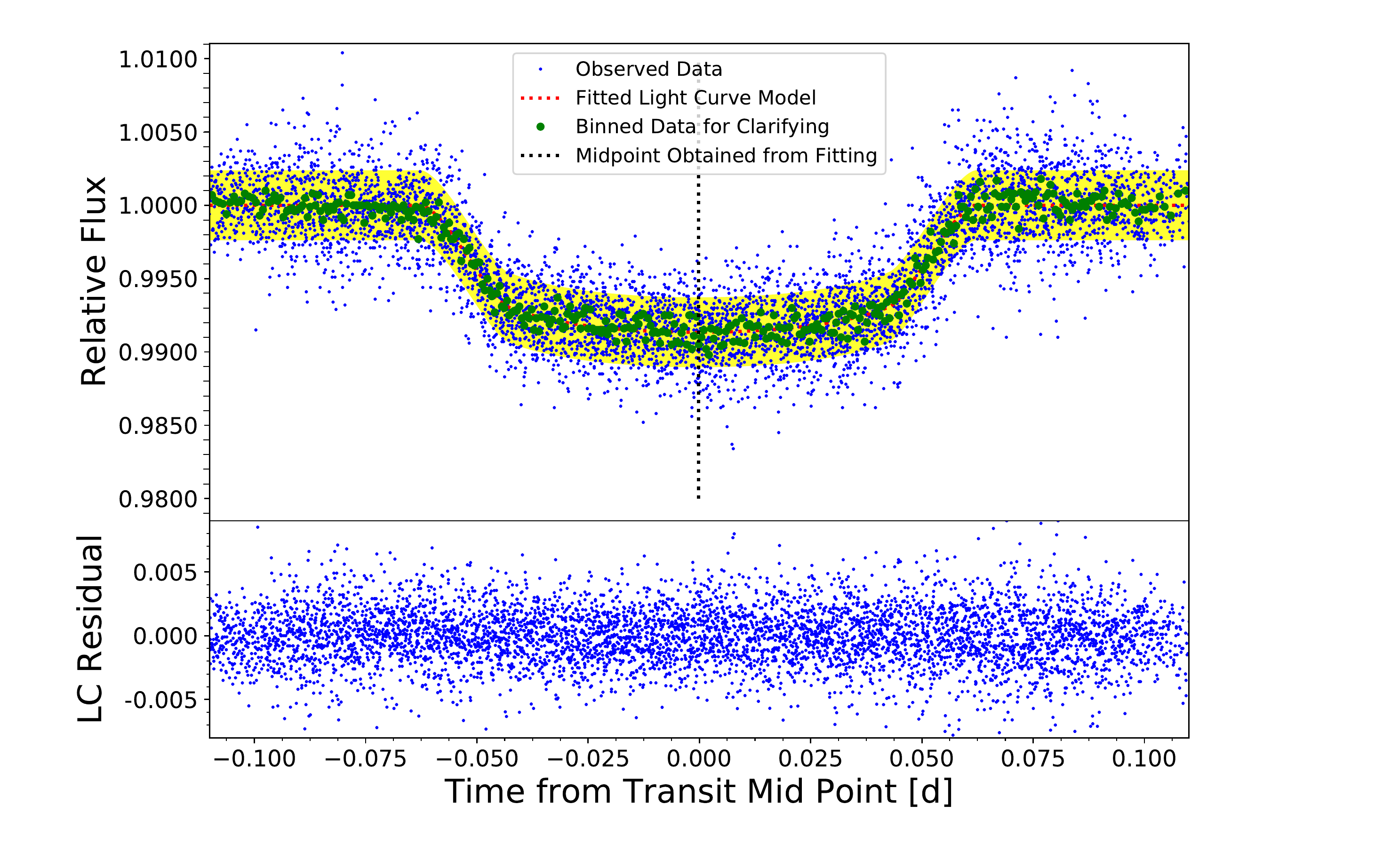}
  \caption{Fits to XO-3b light curve with archival data from \citet{Winn2008}. The symbols are the same as Figure \ref{image:lc}. The standard deviation of the residual (bottom panel) is 2207ppm.}
  \label{image:archivallc}
\end{figure*}

Archival light curves from \citet{Winn2008} are from different band observations. The transit duration should not change when combining light curves in different bands. We note that combining light curves in different bands might influence some transit parameters, e.g., transit depth, ingress/egress shape. We fit the folded light curve, following the same steps as described in Section \ref{sec:fit} (in Figure \ref{image:archivallc}). We fold the light curve in each band and find that the timing durations show no evident difference to the folded light curve in all bands. We note that combined timing duration uncertainty obtained in each band according to error propagation law is a little bit smaller than the uncertainty obtained in modeling the folded light curve in combining bands. In this work, we cite the fitting uncertainty from the all-band light curve as the parameter uncertainty.

The transit duration of the archival folded light curve \citep{Winn2008} is 0.1230$\pm$0.0014 days. As a comparison, the transit duration of the TESS folded light curve is 0.1225$\pm$0.0009 days. The difference ($\sim$ 1 minute) is well within the observational uncertainties and is significantly smaller than the TTV of $\sim$ 17.6 minutes.

However, the impact factor b for XO-3b is possibly very closed to 1/$\sqrt{2}$ which weakens TDV distinguishing evidence towards precession. \citet{Jordan2008} deduce that the logarithmic derivative
of transit duration has a multiplying factor of $1-\frac{b^2}{1-b^2}$. The b for XO-3b would be 0.70$\pm$0.04 if applying orbital parameters obtained from \citet{Bonomo2017}. The TDV would be far smaller than 1 minute with such an impact factor. We thereby conclude precession as a possible physical origin of the observed TTV. Further determination requires more observations and is beyond the scope of this work.

The possibility of R$\o$mer effect is rare that neither long-term radial velocity changing nor significant evidence of stellar companion is found in the previous investigations \citep[][and references therein]{Johns-Krull2008, Winn2008, Winn2009, Wong2014, Stassun2017, Bonomo2017}. \citet{wasp-4b2020} model the relation between the orbital period derivative and stellar acceleration. The acceleration rate would be 212$\pm$10 m s$^{-1}$ yr$^{-1}$, relating the observed period derivative of XO-3b. \citet{Wong2014} report an RV acceleration rate of 8.40$\pm$9.13 m s$^{-1}$ yr$^{-1}$ by modeling the archival RV measurements in a baseline of more than 2000 days \citep[see technical details in][]{Knutson2014}. This reported acceleration rate is too small to generate a comparable TTV observed.

It is known that WASP-12b near Roche limit loses mass at a rate about $10^{-7} M_J$ per year due to the tidal heating in the planet \citep{Li2010} or the interaction of the stellar wind with the planetary magnetosphere \citep{Lai2010}. Such a large mass loss rate probably has an appreciable impact on the orbital decay of WASP-12b. However, for XO-3b, its Roche limit is estimated to be  $\sim 0.4 R_*$ but its periastron to be $\sim 3.4 R_*$, and therefore, the mass loss is unlikely to occur on this planet.

\section{discussion and summary}

We report the evidence of XO-3b TTVs, showing an earlier timing in late 2019 TESS observation at 17.6 minutes when compared to the constant period ephemeris \citep{Wong2014}. A quadratic function, indicating a constant period derivative, is a significantly better fit to timings available than any constant period fit ($\Delta$(BIC) $\ge$ 383). The derived $\dot{P}$ is -6.2$\times$10$^{-9}$$\pm$2.9$\times$10$^{-10}$ days per orbit per day, implying a decay timescale $P$/$\dot{P}$ of 1.4 Myr. We note that a precession model may be also possible though we majorly discuss the tidal dissipation model in explaining observed TTVs.

The theory of equilibrium tide predicts $Q'_\star$ 1.5$\times$10$^{5}$$\pm$6$\times$10$^{3}$ or $Q'_p$ 1.8$\times$10$^{4}$$\pm$8$\times$10$^{2}$, which is consistent with the reported parameters of planet-star system. The TDV should be more significant than TTV which is not the case.

We have scheduled new transit observations in Sitian Early commissioning, using three 30 cm prototype telescopes \citep{SiTian}. The new transit timings would provide further information to discriminate the difference between precession and period decay models.

\begin{acknowledgements}
This work made use of the NASA Exoplanet Archive \footnote{\url{https://exoplanetarchive.ipac.caltech.edu/index.html}} \citep{ExoplanetArchive} and PyAstronomy\footnote{https://github.com/sczesla/PyAstronomy} \citep{pya}. We would like to thank Bo Zhang for the helpful discussion. Fan Yang acknowledge funding from the Cultivation Project for LAMOST Scientific Payoff and Research Achievement of CAMS-CAS. Xing Wei is supported by National Natural Science Foundation of China (NSFC; No.11872246, 12041301), and the Beijing Natural Science Foundation (No. 1202015).

\end{acknowledgements}

\bibliographystyle{aasjournal}
\bibliography{ref}

\begin{thebibliography}{}
\expandafter\ifx\csname natexlab\endcsname\relax\def\natexlab#1{#1}\fi
\providecommand{\url}[1]{\href{#1}{#1}}
\providecommand{\dodoi}[1]{doi:~\href{http://doi.org/#1}{\nolinkurl{#1}}}
\providecommand{\doeprint}[1]{\href{http://ascl.net/#1}{\nolinkurl{http://ascl.net/#1}}}
\providecommand{\doarXiv}[1]{\href{https://arxiv.org/abs/#1}{\nolinkurl{https://arxiv.org/abs/#1}}}

\bibitem[{{Akeson} {et~al.}(2013){Akeson}, {Chen}, {Ciardi}, {Crane}, {Good},
  {Harbut}, {Jackson}, {Kane}, {Laity}, {Leifer}, {Lynn}, {McElroy}, {Papin},
  {Plavchan}, {Ram{\'\i}rez}, {Rey}, {von Braun}, {Wittman}, {Abajian}, {Ali},
  {Beichman}, {Beekley}, {Berriman}, {Berukoff}, {Bryden}, {Chan}, {Groom},
  {Lau}, {Payne}, {Regelson}, {Saucedo}, {Schmitz}, {Stauffer}, {Wyatt}, \&
  {Zhang}}]{ExoplanetArchive}
{Akeson}, R.~L., {Chen}, X., {Ciardi}, D., {et~al.} 2013, \pasp, 125, 989,
  \dodoi{10.1086/672273}

\bibitem[{{Albrecht} {et~al.}(2012){Albrecht}, {Winn}, {Johnson}, {Howard},
  {Marcy}, {Butler}, {Arriagada}, {Crane}, {Shectman}, {Thompson}, {Hirano},
  {Bakos}, \& {Hartman}}]{2012wasp12b}
{Albrecht}, S., {Winn}, J.~N., {Johnson}, J.~A., {et~al.} 2012, \apj, 757, 18,
  \dodoi{10.1088/0004-637X/757/1/18}

\bibitem[{{Antoniciello} {et~al.}(2021){Antoniciello}, {Borsato}, {Lacedelli},
  {Nascimbeni}, {Barrag{\'a}n}, \& {Claudi}}]{Nascimbeni2021}
{Antoniciello}, G., {Borsato}, L., {Lacedelli}, G., {et~al.} 2021, \mnras, 505,
  1567, \dodoi{10.1093/mnras/stab1336}

\bibitem[{{Bodenheimer} {et~al.}(2001){Bodenheimer}, {Lin}, \&
  {Mardling}}]{Bodenheimer2001}
{Bodenheimer}, P., {Lin}, D.~N.~C., \& {Mardling}, R.~A. 2001, \apj, 548, 466,
  \dodoi{10.1086/318667}

\bibitem[{{Bonomo} {et~al.}(2017){Bonomo}, {Desidera}, {Benatti}, {Borsa},
  {Crespi}, {Damasso}, {Lanza}, {Sozzetti}, {Lodato}, {Marzari}, {Boccato},
  {Claudi}, {Cosentino}, {Covino}, {Gratton}, {Maggio}, {Micela}, {Molinari},
  {Pagano}, {Piotto}, {Poretti}, {Smareglia}, {Affer}, {Biazzo}, {Bignamini},
  {Esposito}, {Giacobbe}, {H{\'e}brard}, {Malavolta}, {Maldonado}, {Mancini},
  {Martinez Fiorenzano}, {Masiero}, {Nascimbeni}, {Pedani}, {Rainer}, \&
  {Scandariato}}]{Bonomo2017}
{Bonomo}, A.~S., {Desidera}, S., {Benatti}, S., {et~al.} 2017, \aap, 602, A107,
  \dodoi{10.1051/0004-6361/201629882}

\bibitem[{{Bouma} {et~al.}(2020){Bouma}, {Winn}, {Howard}, {Howell},
  {Isaacson}, {Knutson}, \& {Matson}}]{wasp-4b2020}
{Bouma}, L.~G., {Winn}, J.~N., {Howard}, A.~W., {et~al.} 2020, \apjl, 893, L29,
  \dodoi{10.3847/2041-8213/ab8563}

\bibitem[{{Bouma} {et~al.}(2019){Bouma}, {Winn}, {Baxter}, {Bhatti}, {Dai},
  {Daylan}, {D{\'e}sert}, {Hill}, {Kane}, {Stassun}, {Villasenor}, {Ricker},
  {Vanderspek}, {Latham}, {Seager}, {Jenkins}, {Berta-Thompson}, {Col{\'o}n},
  {Fausnaugh}, {Glidden}, {Guerrero}, {Rodriguez}, {Twicken}, \&
  {Wohler}}]{WASP-4b}
{Bouma}, L.~G., {Winn}, J.~N., {Baxter}, C., {et~al.} 2019, \aj, 157, 217,
  \dodoi{10.3847/1538-3881/ab189f}

\bibitem[{{Campo} {et~al.}(2011){Campo}, {Harrington}, {Hardy}, {Stevenson},
  {Nymeyer}, {Ragozzine}, {Lust}, {Anderson}, {Collier-Cameron}, {Blecic},
  {Britt}, {Bowman}, {Wheatley}, {Loredo}, {Deming}, {Hebb}, {Hellier},
  {Maxted}, {Pollaco}, \& {West}}]{2011wasp12b}
{Campo}, C.~J., {Harrington}, J., {Hardy}, R.~A., {et~al.} 2011, \apj, 727,
  125, \dodoi{10.1088/0004-637X/727/2/125}

\bibitem[{{Claret}(2018)}]{TESSLD}
{Claret}, A. 2018, \aap, 618, A20, \dodoi{10.1051/0004-6361/201833060}

\bibitem[{{Czesla} {et~al.}(2019){Czesla}, {Schr{\"o}ter}, {Schneider},
  {Huber}, {Pfeifer}, {Andreasen}, \& {Zechmeister}}]{pya}
{Czesla}, S., {Schr{\"o}ter}, S., {Schneider}, C.~P., {et~al.} 2019, {PyA:
  Python astronomy-related packages}.
\newblock \doeprint{1906.010}

\bibitem[{{Dawson} \& {Johnson}(2018)}]{Dawson2018}
{Dawson}, R.~I., \& {Johnson}, J.~A. 2018, \araa, 56, 175,
  \dodoi{10.1146/annurev-astro-081817-051853}

\bibitem[{{Deming} {et~al.}(2009){Deming}, {Seager}, {Winn}, {Miller-Ricci},
  {Clampin}, {Lindler}, {Greene}, {Charbonneau}, {Laughlin}, {Ricker},
  {Latham}, \& {Ennico}}]{Deming2009}
{Deming}, D., {Seager}, S., {Winn}, J., {et~al.} 2009, \pasp, 121, 952,
  \dodoi{10.1086/605913}

\bibitem[{{Endl} {et~al.}(2014){Endl}, {Caldwell}, {Barclay}, {Huber},
  {Isaacson}, {Buchhave}, {Brugamyer}, {Robertson}, {Cochran}, {MacQueen},
  {Havel}, {Lucas}, {Howell}, {Fischer}, {Quintana}, \& {Ciardi}}]{Endl2014}
{Endl}, M., {Caldwell}, D.~A., {Barclay}, T., {et~al.} 2014, \apj, 795, 151,
  \dodoi{10.1088/0004-637X/795/2/151}

\bibitem[{{Fabrycky} \& {Tremaine}(2007)}]{Fabrycky2007}
{Fabrycky}, D., \& {Tremaine}, S. 2007, \apj, 669, 1298, \dodoi{10.1086/521702}

\bibitem[{{Fausnaugh} {et~al.}(2021){Fausnaugh}, {Morgan}, {Vanderspek},
  {Pepper}, {Burke}, {Levine}, {Rudat}, {Villase{\~n}or}, {Vezie}, {Goeke},
  {Ricker}, {Latham}, {Seager}, {Winn}, {Jenkins}, {Bakos}, {Barclay},
  {Berta-Thompson}, {Bouma}, {Boyd}, {Brasseur}, {Burt}, {Caldwell},
  {Charbonneau}, {Christensen-Dalsgaard}, {Clampin}, {Collins}, {Col{\'o}n},
  {De Lee}, {Dunham}, {Fleming}, {Fong}, {Garcia Soto}, {Scott Gaudi},
  {Guerrero}, {Hesse}, {Holman}, {Huang}, {Kaltenegger}, {Lissauer},
  {McDermott}, {McLean}, {Mireles}, {Mullally}, {Oelkers}, {Paegert},
  {P{\'a}l}, {Quintana}, {Rinehart}, {Rodriguez}, {Rose}, {Sasselov},
  {Schlieder}, {Sha}, {Shporer}, {Smith}, {Stassun}, {Tenenbaum}, {Ting},
  {Torres}, {Twicken}, {Vanderburg}, {Wohler}, \& {Yu}}]{Fausnaugh2021}
{Fausnaugh}, M., {Morgan}, E., {Vanderspek}, R., {et~al.} 2021, \pasp, 133,
  095002, \dodoi{10.1088/1538-3873/ac1d3f}

\bibitem[{{Feinstein} {et~al.}(2019){Feinstein}, {Montet}, {Foreman-Mackey},
  {Bedell}, {Saunders}, {Bean}, {Christiansen}, {Hedges}, {Luger}, {Scolnic},
  \& {Cardoso}}]{Feinstein2019}
{Feinstein}, A.~D., {Montet}, B.~T., {Foreman-Mackey}, D., {et~al.} 2019,
  \pasp, 131, 094502, \dodoi{10.1088/1538-3873/ab291c}

\bibitem[{{Foreman-Mackey} {et~al.}(2013){Foreman-Mackey}, {Hogg}, {Lang}, \&
  {Goodman}}]{emcee}
{Foreman-Mackey}, D., {Hogg}, D.~W., {Lang}, D., \& {Goodman}, J. 2013, \pasp,
  125, 306, \dodoi{10.1086/670067}

\bibitem[{{Gaia Collaboration} {et~al.}(2018){Gaia Collaboration}, {Brown},
  {Vallenari}, {Prusti}, {de Bruijne}, {Babusiaux}, {Bailer-Jones}, {Biermann},
  {Evans}, {Eyer}, {Jansen}, {Jordi}, {Klioner}, {Lammers}, {Lindegren},
  {Luri}, {Mignard}, {Panem}, {Pourbaix}, {Randich}, {Sartoretti}, {Siddiqui},
  {Soubiran}, {van Leeuwen}, {Walton}, {Arenou}, {Bastian}, {Cropper},
  {Drimmel}, {Katz}, {Lattanzi}, {Bakker}, {Cacciari}, {Casta{\~n}eda},
  {Chaoul}, {Cheek}, {De Angeli}, {Fabricius}, {Guerra}, {Holl}, {Masana},
  {Messineo}, {Mowlavi}, {Nienartowicz}, {Panuzzo}, {Portell}, {Riello},
  {Seabroke}, {Tanga}, {Th{\'e}venin}, {Gracia-Abril}, {Comoretto},
  {Garcia-Reinaldos}, {Teyssier}, {Altmann}, {Andrae}, {Audard},
  {Bellas-Velidis}, {Benson}, {Berthier}, {Blomme}, {Burgess}, {Busso},
  {Carry}, {Cellino}, {Clementini}, {Clotet}, {Creevey}, {Davidson}, {De
  Ridder}, {Delchambre}, {Dell'Oro}, {Ducourant},
  {Fern{\'a}ndez-Hern{\'a}ndez}, {Fouesneau}, {Fr{\'e}mat}, {Galluccio},
  {Garc{\'\i}a-Torres}, {Gonz{\'a}lez-N{\'u}{\~n}ez}, {Gonz{\'a}lez-Vidal},
  {Gosset}, {Guy}, {Halbwachs}, {Hambly}, {Harrison}, {Hern{\'a}ndez},
  {Hestroffer}, {Hodgkin}, {Hutton}, {Jasniewicz}, {Jean-Antoine-Piccolo},
  {Jordan}, {Korn}, {Krone-Martins}, {Lanzafame}, {Lebzelter}, {L{\"o}ffler},
  {Manteiga}, {Marrese}, {Mart{\'\i}n-Fleitas}, {Moitinho}, {Mora}, {Muinonen},
  {Osinde}, {Pancino}, {Pauwels}, {Petit}, {Recio-Blanco}, {Richards},
  {Rimoldini}, {Robin}, {Sarro}, {Siopis}, {Smith}, {Sozzetti}, {S{\"u}veges},
  {Torra}, {van Reeven}, {Abbas}, {Abreu Aramburu}, {Accart}, {Aerts},
  {Altavilla}, {{\'A}lvarez}, {Alvarez}, {Alves}, {Anderson}, {Andrei},
  {Anglada Varela}, {Antiche}, {Antoja}, {Arcay}, {Astraatmadja}, {Bach},
  {Baker}, {Balaguer-N{\'u}{\~n}ez}, {Balm}, {Barache}, {Barata}, {Barbato},
  {Barblan}, {Barklem}, {Barrado}, {Barros}, {Barstow}, {Bartholom{\'e}
  Mu{\~n}oz}, {Bassilana}, {Becciani}, {Bellazzini}, {Berihuete}, {Bertone},
  {Bianchi}, {Bienaym{\'e}}, {Blanco-Cuaresma}, {Boch}, {Boeche}, {Bombrun},
  {Borrachero}, {Bossini}, {Bouquillon}, {Bourda}, {Bragaglia}, {Bramante},
  {Breddels}, {Bressan}, {Brouillet}, {Br{\"u}semeister}, {Brugaletta},
  {Bucciarelli}, {Burlacu}, {Busonero}, {Butkevich}, {Buzzi}, {Caffau},
  {Cancelliere}, {Cannizzaro}, {Cantat-Gaudin}, {Carballo}, {Carlucci},
  {Carrasco}, {Casamiquela}, {Castellani}, {Castro-Ginard}, {Charlot},
  {Chemin}, {Chiavassa}, {Cocozza}, {Costigan}, {Cowell}, {Crifo}, {Crosta},
  {Crowley}, {Cuypers}, {Dafonte}, {Damerdji}, {Dapergolas}, {David}, {David},
  {de Laverny}, {De Luise}, {De March}, {de Martino}, {de Souza}, {de Torres},
  {Debosscher}, {del Pozo}, {Delbo}, {Delgado}, {Delgado}, {Di Matteo},
  {Diakite}, {Diener}, {Distefano}, {Dolding}, {Drazinos}, {Dur{\'a}n},
  {Edvardsson}, {Enke}, {Eriksson}, {Esquej}, {Eynard Bontemps}, {Fabre},
  {Fabrizio}, {Faigler}, {Falc{\~a}o}, {Farr{\`a}s Casas}, {Federici},
  {Fedorets}, {Fernique}, {Figueras}, {Filippi}, {Findeisen}, {Fonti},
  {Fraile}, {Fraser}, {Fr{\'e}zouls}, {Gai}, {Galleti}, {Garabato},
  {Garc{\'\i}a-Sedano}, {Garofalo}, {Garralda}, {Gavel}, {Gavras}, {Gerssen},
  {Geyer}, {Giacobbe}, {Gilmore}, {Girona}, {Giuffrida}, {Glass}, {Gomes},
  {Granvik}, {Gueguen}, {Guerrier}, {Guiraud}, {Guti{\'e}rrez-S{\'a}nchez},
  {Haigron}, {Hatzidimitriou}, {Hauser}, {Haywood}, {Heiter}, {Helmi}, {Heu},
  {Hilger}, {Hobbs}, {Hofmann}, {Holland}, {Huckle}, {Hypki}, {Icardi},
  {Jan{\ss}en}, {Jevardat de Fombelle}, {Jonker}, {Juh{\'a}sz}, {Julbe},
  {Karampelas}, {Kewley}, {Klar}, {Kochoska}, {Kohley}, {Kolenberg},
  {Kontizas}, {Kontizas}, {Koposov}, {Kordopatis}, {Kostrzewa-Rutkowska},
  {Koubsky}, {Lambert}, {Lanza}, {Lasne}, {Lavigne}, {Le Fustec}, {Le
  Poncin-Lafitte}, {Lebreton}, {Leccia}, {Leclerc}, {Lecoeur-Taibi},
  {Lenhardt}, {Leroux}, {Liao}, {Licata}, {Lindstr{\o}m}, {Lister}, {Livanou},
  {Lobel}, {L{\'o}pez}, {Managau}, {Mann}, {Mantelet}, {Marchal}, {Marchant},
  {Marconi}, {Marinoni}, {Marschalk{\'o}}, {Marshall}, {Martino}, {Marton},
  {Mary}, {Massari}, {Matijevi{\v{c}}}, {Mazeh}, {McMillan}, {Messina},
  {Michalik}, {Millar}, {Molina}, {Molinaro}, {Moln{\'a}r}, {Montegriffo},
  {Mor}, {Morbidelli}, {Morel}, {Morris}, {Mulone}, {Muraveva}, {Musella},
  {Nelemans}, {Nicastro}, {Noval}, {O'Mullane}, {Ord{\'e}novic},
  {Ord{\'o}{\~n}ez-Blanco}, {Osborne}, {Pagani}, {Pagano}, {Pailler},
  {Palacin}, {Palaversa}, {Panahi}, {Pawlak}, {Piersimoni}, {Pineau}, {Plachy},
  {Plum}, {Poggio}, {Poujoulet}, {Pr{\v{s}}a}, {Pulone}, {Racero}, {Ragaini},
  {Rambaux}, {Ramos-Lerate}, {Regibo}, {Reyl{\'e}}, {Riclet}, {Ripepi}, {Riva},
  {Rivard}, {Rixon}, {Roegiers}, {Roelens}, {Romero-G{\'o}mez}, {Rowell},
  {Royer}, {Ruiz-Dern}, {Sadowski}, {Sagrist{\`a} Sell{\'e}s}, {Sahlmann},
  {Salgado}, {Salguero}, {Sanna}, {Santana-Ros}, {Sarasso}, {Savietto},
  {Schultheis}, {Sciacca}, {Segol}, {Segovia}, {S{\'e}gransan}, {Shih},
  {Siltala}, {Silva}, {Smart}, {Smith}, {Solano}, {Solitro}, {Sordo}, {Soria
  Nieto}, {Souchay}, {Spagna}, {Spoto}, {Stampa}, {Steele},
  {Steidelm{\"u}ller}, {Stephenson}, {Stoev}, {Suess}, {Surdej}, {Szabados},
  {Szegedi-Elek}, {Tapiador}, {Taris}, {Tauran}, {Taylor}, {Teixeira},
  {Terrett}, {Teyssandier}, {Thuillot}, {Titarenko}, {Torra Clotet}, {Turon},
  {Ulla}, {Utrilla}, {Uzzi}, {Vaillant}, {Valentini}, {Valette}, {van Elteren},
  {Van Hemelryck}, {van Leeuwen}, {Vaschetto}, {Vecchiato}, {Veljanoski},
  {Viala}, {Vicente}, {Vogt}, {von Essen}, {Voss}, {Votruba}, {Voutsinas},
  {Walmsley}, {Weiler}, {Wertz}, {Wevers}, {Wyrzykowski}, {Yoldas},
  {{\v{Z}}erjal}, {Ziaeepour}, {Zorec}, {Zschocke}, {Zucker}, {Zurbach}, \&
  {Zwitter}}]{GaiaDr2}
{Gaia Collaboration}, {Brown}, A.~G.~A., {Vallenari}, A., {et~al.} 2018, \aap,
  616, A1, \dodoi{10.1051/0004-6361/201833051}

\bibitem[{{Goldreich} \& {Soter}(1966)}]{Goldreich1966}
{Goldreich}, P., \& {Soter}, S. 1966, \icarus, 5, 375,
  \dodoi{10.1016/0019-1035(66)90051-0}

\bibitem[{{Guerrero} {et~al.}(2021){Guerrero}, {Seager}, {Huang}, {Vanderburg},
  {Garcia Soto}, {Mireles}, {Hesse}, {Fong}, {Glidden}, {Shporer}, {Latham},
  {Collins}, {Quinn}, {Burt}, {Dragomir}, {Crossfield}, {Vanderspek},
  {Fausnaugh}, {Burke}, {Ricker}, {Daylan}, {Essack}, {G{\"u}nther}, {Osborn},
  {Pepper}, {Rowden}, {Sha}, {Villanueva}, {Yahalomi}, {Yu}, {Ballard},
  {Batalha}, {Berardo}, {Chontos}, {Dittmann}, {Esquerdo}, {Mikal-Evans},
  {Jayaraman}, {Krishnamurthy}, {Louie}, {Mehrle}, {Niraula}, {Rackham},
  {Rodriguez}, {Rowden}, {Sousa-Silva}, {Watanabe}, {Wong}, {Zhan},
  {Zivanovic}, {Christiansen}, {Ciardi}, {Swain}, {Lund}, {Mullally},
  {Fleming}, {Rodriguez}, {Boyd}, {Quintana}, {Barclay}, {Col{\'o}n},
  {Rinehart}, {Schlieder}, {Clampin}, {Jenkins}, {Twicken}, {Caldwell},
  {Coughlin}, {Henze}, {Lissauer}, {Morris}, {Rose}, {Smith}, {Tenenbaum},
  {Ting}, {Wohler}, {Bakos}, {Bean}, {Berta-Thompson}, {Bieryla}, {Bouma},
  {Buchhave}, {Butler}, {Charbonneau}, {Doty}, {Ge}, {Holman}, {Howard},
  {Kaltenegger}, {Kane}, {Kjeldsen}, {Kreidberg}, {Lin}, {Minsky}, {Narita},
  {Paegert}, {P{\'a}l}, {Palle}, {Sasselov}, {Spencer}, {Sozzetti}, {Stassun},
  {Torres}, {Udry}, \& {Winn}}]{TOIcatalog}
{Guerrero}, N.~M., {Seager}, S., {Huang}, C.~X., {et~al.} 2021, \apjs, 254, 39,
  \dodoi{10.3847/1538-4365/abefe1}

\bibitem[{{Hansen}(2010)}]{Hansen2010}
{Hansen}, B. M.~S. 2010, \apj, 723, 285, \dodoi{10.1088/0004-637X/723/1/285}

\bibitem[{{H{\'e}brard} {et~al.}(2008){H{\'e}brard}, {Bouchy}, {Pont},
  {Loeillet}, {Rabus}, {Bonfils}, {Moutou}, {Boisse}, {Delfosse}, {Desort},
  {Eggenberger}, {Ehrenreich}, {Forveille}, {Lagrange}, {Lovis}, {Mayor},
  {Pepe}, {Perrier}, {Queloz}, {Santos}, {S{\'e}gransan}, {Udry}, \&
  {Vidal-Madjar}}]{Hebrard2008}
{H{\'e}brard}, G., {Bouchy}, F., {Pont}, F., {et~al.} 2008, \aap, 488, 763,
  \dodoi{10.1051/0004-6361:200810056}

\bibitem[{{Hut}(1981)}]{Hut1981}
{Hut}, P. 1981, \aap, 99, 126

\bibitem[{{Johns-Krull} {et~al.}(2008){Johns-Krull}, {McCullough}, {Burke},
  {Valenti}, {Janes}, {Heasley}, {Prato}, {Bissinger}, {Fleenor}, {Foote},
  {Garcia-Melendo}, {Gary}, {Howell}, {Mallia}, {Masi}, \&
  {Vanmunster}}]{Johns-Krull2008}
{Johns-Krull}, C.~M., {McCullough}, P.~R., {Burke}, C.~J., {et~al.} 2008, \apj,
  677, 657, \dodoi{10.1086/528950}

\bibitem[{{Jord{\'a}n} \& {Bakos}(2008)}]{Jordan2008}
{Jord{\'a}n}, A., \& {Bakos}, G.~{\'A}. 2008, \apj, 685, 543,
  \dodoi{10.1086/590549}

\bibitem[{{Juri{\'c}} \& {Tremaine}(2008)}]{Juric2008}
{Juri{\'c}}, M., \& {Tremaine}, S. 2008, \apj, 686, 603, \dodoi{10.1086/590047}

\bibitem[{{Kane} {et~al.}(2021){Kane}, {Bean}, {Campante}, {Dalba},
  {Fetherolf}, {Mocnik}, {Ostberg}, {Pepper}, {Simpson}, {Turnbull}, {Ricker},
  {Vanderspek}, {Latham}, {Seager}, {Winn}, {Jenkins}, {Huber}, \&
  {Chaplin}}]{Kane2021}
{Kane}, S.~R., {Bean}, J.~L., {Campante}, T.~L., {et~al.} 2021, \pasp, 133,
  014402, \dodoi{10.1088/1538-3873/abc610}

\bibitem[{Kass \& Raftery(1995)}]{BICmostcited}
Kass, R.~E., \& Raftery, A.~E. 1995, Journal of the American Statistical
  Association, 90, 773, \dodoi{10.1080/01621459.1995.10476572}

\bibitem[{{Kempton} {et~al.}(2018){Kempton}, {Bean}, {Louie}, {Deming}, {Koll},
  {Mansfield}, {Christiansen}, {L{\'o}pez-Morales}, {Swain}, {Zellem},
  {Ballard}, {Barclay}, {Barstow}, {Batalha}, {Beatty}, {Berta-Thompson},
  {Birkby}, {Buchhave}, {Charbonneau}, {Cowan}, {Crossfield}, {de Val-Borro},
  {Doyon}, {Dragomir}, {Gaidos}, {Heng}, {Hu}, {Kane}, {Kreidberg}, {Mallonn},
  {Morley}, {Narita}, {Nascimbeni}, {Pall{\'e}}, {Quintana}, {Rauscher},
  {Seager}, {Shkolnik}, {Sing}, {Sozzetti}, {Stassun}, {Valenti}, \& {von
  Essen}}]{Kempton2018}
{Kempton}, E. M.~R., {Bean}, J.~L., {Louie}, D.~R., {et~al.} 2018, \pasp, 130,
  114401, \dodoi{10.1088/1538-3873/aadf6f}

\bibitem[{{Knutson} {et~al.}(2014){Knutson}, {Fulton}, {Montet}, {Kao}, {Ngo},
  {Howard}, {Crepp}, {Hinkley}, {Bakos}, {Batygin}, {Johnson}, {Morton}, \&
  {Muirhead}}]{Knutson2014}
{Knutson}, H.~A., {Fulton}, B.~J., {Montet}, B.~T., {et~al.} 2014, \apj, 785,
  126, \dodoi{10.1088/0004-637X/785/2/126}

\bibitem[{{Lai} {et~al.}(2010){Lai}, {Helling}, \& {van den Heuvel}}]{Lai2010}
{Lai}, D., {Helling}, C., \& {van den Heuvel}, E.~P.~J. 2010, \apj, 721, 923,
  \dodoi{10.1088/0004-637X/721/2/923}

\bibitem[{{Leconte} {et~al.}(2010){Leconte}, {Chabrier}, {Baraffe}, \&
  {Levrard}}]{Themodel}
{Leconte}, J., {Chabrier}, G., {Baraffe}, I., \& {Levrard}, B. 2010, \aap, 516,
  A64, \dodoi{10.1051/0004-6361/201014337}

\bibitem[{{Levrard} {et~al.}(2009){Levrard}, {Winisdoerffer}, \&
  {Chabrier}}]{Levrard2009}
{Levrard}, B., {Winisdoerffer}, C., \& {Chabrier}, G. 2009, \apjl, 692, L9,
  \dodoi{10.1088/0004-637X/692/1/L9}

\bibitem[{{Li} {et~al.}(2010){Li}, {Miller}, {Lin}, \& {Fortney}}]{Li2010}
{Li}, S.-L., {Miller}, N., {Lin}, D. N.~C., \& {Fortney}, J.~J. 2010, \nat,
  463, 1054, \dodoi{10.1038/nature08715}

\bibitem[{{Lin} {et~al.}(1996){Lin}, {Bodenheimer}, \& {Richardson}}]{Lin1996}
{Lin}, D.~N.~C., {Bodenheimer}, P., \& {Richardson}, D.~C. 1996, \nat, 380,
  606, \dodoi{10.1038/380606a0}

\bibitem[{{Liu} {et~al.}(2021){Liu}, {Soria}, {Wu}, {Wu}, \& {Shang}}]{SiTian}
{Liu}, J., {Soria}, R., {Wu}, X.-F., {Wu}, H., \& {Shang}, Z. 2021, An. Acad.
  Bras. Ci{\^e}nc. vol.93 supl.1, 93, 20200628,
  \dodoi{10.1590/0001-3765202120200628}

\bibitem[{{Mandel} \& {Agol}(2002)}]{Mandel_Agol2002}
{Mandel}, K., \& {Agol}, E. 2002, \apj, 580, L171, \dodoi{10.1086/345520}

\bibitem[{{Naoz} {et~al.}(2011){Naoz}, {Farr}, {Lithwick}, {Rasio}, \&
  {Teyssandier}}]{Naoz2011}
{Naoz}, S., {Farr}, W.~M., {Lithwick}, Y., {Rasio}, F.~A., \& {Teyssandier}, J.
  2011, \nat, 473, 187, \dodoi{10.1038/nature10076}

\bibitem[{{Ogilvie}(2014)}]{Ogilvie2014}
{Ogilvie}, G.~I. 2014, Annu. Rev. Astron. Astrophys., 52, 171

\bibitem[{{P{\'a}l} \& {Kocsis}(2008)}]{Pal2008}
{P{\'a}l}, A., \& {Kocsis}, B. 2008, \mnras, 389, 191,
  \dodoi{10.1111/j.1365-2966.2008.13512.x}

\bibitem[{{Pan} {et~al.}(2021){Pan}, {Fu}, {Zhang}, {Wang}, {Zong}, {Li}, \&
  {Zhang}}]{PanFu2021}
{Pan}, Y., {Fu}, J.-N., {Zhang}, X., {et~al.} 2021, \pasp, 133, 044202,
  \dodoi{10.1088/1538-3873/abef77}

\bibitem[{{Patil} {et~al.}(2010){Patil}, {Huard}, \& {Fonnesbeck}}]{pymc}
{Patil}, A., {Huard}, D., \& {Fonnesbeck}, C.~J. 2010, Journal of Statistical
  Software, 35, 1, \dodoi{10.18637/jss.v035.i04}

\bibitem[{{Patra} {et~al.}(2017){Patra}, {Winn}, {Holman}, {Yu}, {Deming}, \&
  {Dai}}]{2017wasp12b}
{Patra}, K.~C., {Winn}, J.~N., {Holman}, M.~J., {et~al.} 2017, \aj, 154, 4,
  \dodoi{10.3847/1538-3881/aa6d75}

\bibitem[{{Penev} {et~al.}(2014){Penev}, {Zhang}, \& {Jackson}}]{Penev2014}
{Penev}, K., {Zhang}, M., \& {Jackson}, B. 2014, \pasp, 126, 553,
  \dodoi{10.1086/677042}

\bibitem[{{Petrovich}(2015)}]{Petrovich2015}
{Petrovich}, C. 2015, \apj, 805, 75, \dodoi{10.1088/0004-637X/805/1/75}

\bibitem[{{Placek} {et~al.}(2016){Placek}, {Knuth}, \&
  {Angerhausen}}]{Placek2016}
{Placek}, B., {Knuth}, K.~H., \& {Angerhausen}, D. 2016, \pasp, 128, 074503,
  \dodoi{10.1088/1538-3873/128/965/074503}

\bibitem[{{Ragozzine} \& {Wolf}(2009)}]{Ragozzine2009}
{Ragozzine}, D., \& {Wolf}, A.~S. 2009, \apj, 698, 1778,
  \dodoi{10.1088/0004-637X/698/2/1778}

\bibitem[{{Ricker} {et~al.}(2015){Ricker}, {Winn}, {Vanderspek}, {Latham},
  {Bakos}, {Bean}, {Berta-Thompson}, {Brown}, {Buchhave}, \&
  {Butler}}]{Ricker2015}
{Ricker}, G.~R., {Winn}, J.~N., {Vanderspek}, R., {et~al.} 2015, Journal of
  Astronomical Telescopes, Instruments, and Systems, 1, 014003,
  \dodoi{10.1117/1.JATIS.1.1.014003}

\bibitem[{Shan {et~al.}(2021)Shan, Yang, Lu, Wei, Tian, Zhang, Guo, Cui, Yang,
  Zhang, \& Liu}]{shan2021}
Shan, S.-S., Yang, F., Lu, Y.-J., {et~al.} 2021.
\newblock \doarXiv{2111.06678}

\bibitem[{{Shara} {et~al.}(2016){Shara}, {Hurley}, \& {Mardling}}]{Shara2016}
{Shara}, M.~M., {Hurley}, J.~R., \& {Mardling}, R.~A. 2016, \apj, 816, 59,
  \dodoi{10.3847/0004-637X/816/2/59}

\bibitem[{{Stassun} {et~al.}(2017){Stassun}, {Collins}, \&
  {Gaudi}}]{Stassun2017}
{Stassun}, K.~G., {Collins}, K.~A., \& {Gaudi}, B.~S. 2017, \aj, 153, 136,
  \dodoi{10.3847/1538-3881/aa5df3}

\bibitem[{{Stassun} {et~al.}(2019){Stassun}, {Oelkers}, {Paegert}, {Torres},
  {Pepper}, {De Lee}, {Collins}, {Latham}, {Muirhead}, {Chittidi},
  {Rojas-Ayala}, {Fleming}, {Rose}, {Tenenbaum}, {Ting}, {Kane}, {Barclay},
  {Bean}, {Brassuer}, {Charbonneau}, {Ge}, {Lissauer}, {Mann}, {McLean},
  {Mullally}, {Narita}, {Plavchan}, {Ricker}, {Sasselov}, {Seager}, {Sharma},
  {Shiao}, {Sozzetti}, {Stello}, {Vanderspek}, {Wallace}, \&
  {Winn}}]{Stassun2019}
{Stassun}, K.~G., {Oelkers}, R.~J., {Paegert}, M., {et~al.} 2019, \aj, 158,
  138, \dodoi{10.3847/1538-3881/ab3467}

\bibitem[{{Stumpe} {et~al.}(2014){Stumpe}, {Smith}, {Catanzarite}, {Van Cleve},
  {Jenkins}, {Twicken}, \& {Girouard}}]{Stumpe2014}
{Stumpe}, M.~C., {Smith}, J.~C., {Catanzarite}, J.~H., {et~al.} 2014, \pasp,
  126, 100, \dodoi{10.1086/674989}

\bibitem[{{Turner} {et~al.}(2021){Turner}, {Ridden-Harper}, \&
  {Jayawardhana}}]{2021wasp12bTESS}
{Turner}, J.~D., {Ridden-Harper}, A., \& {Jayawardhana}, R. 2021, \aj, 161, 72,
  \dodoi{10.3847/1538-3881/abd178}

\bibitem[{{Winn} {et~al.}(2008){Winn}, {Holman}, {Torres}, {McCullough},
  {Johns-Krull}, {Latham}, {Shporer}, {Mazeh}, {Garcia-Melendo}, {Foote},
  {Esquerdo}, \& {Everett}}]{Winn2008}
{Winn}, J.~N., {Holman}, M.~J., {Torres}, G., {et~al.} 2008, \apj, 683, 1076,
  \dodoi{10.1086/589737}

\bibitem[{{Winn} {et~al.}(2009){Winn}, {Johnson}, {Fabrycky}, {Howard},
  {Marcy}, {Narita}, {Crossfield}, {Suto}, {Turner}, {Esquerdo}, \&
  {Holman}}]{Winn2009}
{Winn}, J.~N., {Johnson}, J.~A., {Fabrycky}, D., {et~al.} 2009, \apj, 700, 302,
  \dodoi{10.1088/0004-637X/700/1/302}

\bibitem[{{Wong} {et~al.}(2014){Wong}, {Knutson}, {Cowan}, {Lewis}, {Agol},
  {Burrows}, {Deming}, {Fortney}, {Fulton}, {Langton}, {Laughlin}, \&
  {Showman}}]{Wong2014}
{Wong}, I., {Knutson}, H.~A., {Cowan}, N.~B., {et~al.} 2014, \apj, 794, 134,
  \dodoi{10.1088/0004-637X/794/2/134}

\bibitem[{{Wu} \& {Lithwick}(2011)}]{Wu2011}
{Wu}, Y., \& {Lithwick}, Y. 2011, \apj, 735, 109,
  \dodoi{10.1088/0004-637X/735/2/109}

\bibitem[{{Wu} \& {Murray}(2003)}]{Wu2003}
{Wu}, Y., \& {Murray}, N. 2003, \apj, 589, 605, \dodoi{10.1086/374598}

\bibitem[{{Yang} {et~al.}(2022){Yang}, {Chary}, \& {Liu}}]{Yangatmos}
{Yang}, F., {Chary}, R.-R., \& {Liu}, J.-F. 2022, \aj, 163, 42,
  \dodoi{10.3847/1538-3881/ac3b4e}

\bibitem[{Yang {et~al.}(2021)Yang, Shan, Guo, Wei, Zhang, Ji, Gao, \&
  Liu}]{Yanghats5b}
Yang, F., Shan, S.-S., Guo, R., {et~al.} 2021, Astrophysics and Space Science,
  366, 83, \dodoi{10.1007/s10509-021-03989-5}

\bibitem[{{Yang} {et~al.}(2020){Yang}, {Long}, {Shan}, {Zhang}, {Guo}, {Bai},
  {Bai}, {Cui}, {Wang}, \& {Liu}}]{Yang2020}
{Yang}, F., {Long}, R.~J., {Shan}, S.-S., {et~al.} 2020, \apjs, 249, 31,
  \dodoi{10.3847/1538-4365/ab9b77}

\bibitem[{{Yang} {et~al.}(2021){Yang}, {Long}, {Liu}, {Shan}, {Guo}, {Zhang},
  {Yi}, {Zheng}, \& {Zhao}}]{YangLD}
{Yang}, F., {Long}, R.~J., {Liu}, J.-f., {et~al.} 2021, \aj, 161, 294,
  \dodoi{10.3847/1538-3881/abf92f}

\bibitem[{{Yee} {et~al.}(2020){Yee}, {Winn}, {Knutson}, {Patra},
  {Vissapragada}, {Zhang}, {Holman}, {Shporer}, \& {Wright}}]{2020wasp12b}
{Yee}, S.~W., {Winn}, J.~N., {Knutson}, H.~A., {et~al.} 2020, \apjl, 888, L5,
  \dodoi{10.3847/2041-8213/ab5c16}

\end{thebibliography}
\end{document}